\documentclass{JHEP3}

\pdfoutput=1

\usepackage{graphicx}
\usepackage{amssymb}
\usepackage{slashed}
\usepackage{amssymb}
\usepackage{amsmath}
\usepackage{url}

\newcommand{\bi}{\begin{itemize}}
\newcommand{\ei}{\end{itemize}}
\newcommand{\be}{\begin{equation}}
\newcommand{\ee}{\end{equation}}

\DeclareMathOperator{\tr}{Tr}

\newcommand{\gsim}{\gtrsim}

\title{New Physics Signals in Longitudinal Gauge Boson Scattering at the LHC}
\author{Tao Han$^a$, David Krohn$^b$, Lian-Tao Wang$^b$, and Wenhan Zhu$^b$ \\ 
$^a$Department of Physics, University of Wisconsin, Madison, WI 53706  \\
 $^b$Department of Physics, Princeton University, Princeton, NJ 08544  \\ \email{than@hep.wisc.edu,dkrohn@princeton.edu,lianwang@princeton.edu,wenhanz@princeton.edu} }

\abstract{We introduce a novel technique designed to look for signatures of new physics in
 vector boson fusion processes at the TeV scale.  This functions by measuring the polarization of the 
 vector bosons to determine the {\it relative} longitudinal to transverse production.  In studying 
 this ratio we can directly probe the high energy $E^2$-growth of longitudinal vector boson scattering 
 amplitudes characteristic of models with non-Standard Model (SM) interactions.  We will focus on 
 studying models parameterized by an effective Lagrangian that include a light Higgs with 
 non-SM couplings arising from TeV scale new physics associated with the electroweak symmetry breaking, 
 although our technique can be used in more general scenarios.  We will show that this technique is 
 stable against the large uncertainties that can result from  variations in the factorization scale, improving
 upon previous studies that measure cross section alone.  }


\preprint{MADPH--09-1544}

\begin{document}

\section{Vector Boson Fusion as a Probe of New Physics}
The Large Hadron Collider (LHC) was built to elucidate the physics behind electroweak symmetry breaking (EWSB).  In a sense, 
it must succeed in finding some new physics 
because the partial wave amplitudes for $V_LV_L\rightarrow V_LV_L$ scattering,\footnote{By $V_L$ we denote 
a longitudinally polarized electroweak vector boson.} calculated in the absence of a Higgs or other new physics, begin to violate unitarity at the 
TeV scale.  Therefore, either new weakly-coupled light particles must come in to unitarize the amplitudes, or 
we will see new strong interactions in the electroweak sector.

While many models of EWSB have been proposed, precision experiments such as LEP seem to favor a model 
employing a $\mathcal{O}(100)~\rm{GeV}$ scalar with the quantum numbers and approximate couplings of the
Standard Model (SM) Higgs~\cite{:2005ema,Barbieri:2004qk}.  Many models of new physics already include such a particle, 
 oftentimes with couplings deviating slightly from those of the SM, {\it e.g.} little Higgs~\cite{ArkaniHamed:2001nc} 
 and holographic Higgs models~\cite{Contino:2003ve}.  Ideally, such models would be identified and studied 
 at the LHC through the production of their intrinsic new particles.  However, the finite energy reach and large
  backgrounds at the LHC could make discovering any new states very difficult. 

Thus we will focus on these non-SM light Higgs scenarios, both because they are favored by precision data and because they 
are perhaps the most difficult to distinguish from the SM. 
To study these setups we will take a model-independent approach, employing an effective field 
theory to parameterize the effects of new physics~\cite{Buchmuller:1985jz,Leung:1984ni,Wudka:1994ny,Hagiwara:1992eh,Hagiwara:1993ck,Hagiwara:2002fs}.
We will see that the general phenomenology of the Higgs sector is captured by the coefficients of a small number of dimension-6 
operators~\cite{Barger:2003rs,Giudice:2007fh}, only one of which is relevant to the vector boson fusion process we wish to study.

\FIGURE{
\includegraphics[scale=0.7]{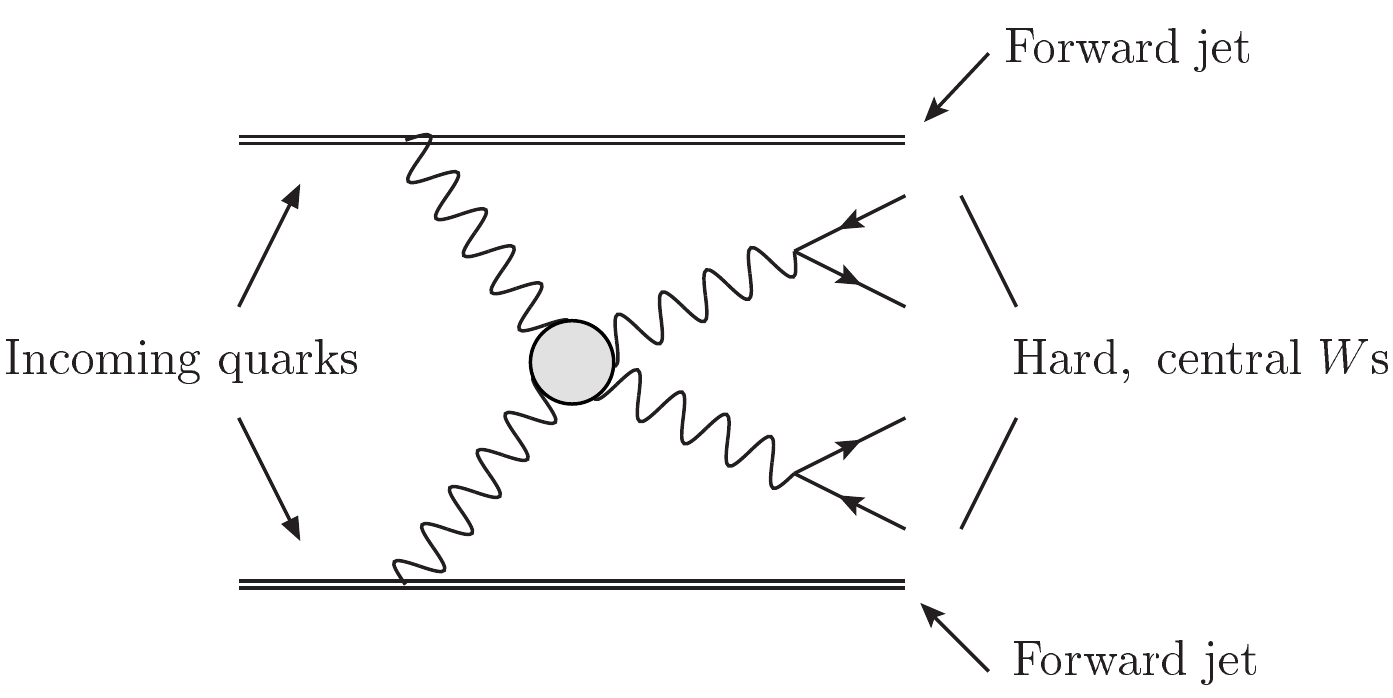}
\caption{
\label{fig:vbf_illus} Illustration for vector boson fusion.}}

Vector boson fusion (VBF)
is the process in which vector bosons radiated by initial state quarks scatter into vector bosons (see Fig.~\ref{fig:vbf_illus}).  
This process is intimately tied to EWSB: just as the pion is a Nambu-Goldstone boson (NGB) and $\pi \pi$ scattering 
can be used to understand chiral symmetry breaking, at high energies longitudinally 
polarized vector bosons take on the behavior of the NGBs from EWSB. 
In the absence of a Higgs boson or other new physics responsible for the EWSB,  the scattering amplitudes probed 
by VBF would violate perturbative unitarity~\cite{Dicus:1992vj,Lee:1977eg,Veltman:1976rt,Chanowitz:1985hj} at around 
$1~\rm{TeV}$ (see the discussion in appendix~\ref{sec:scatamp}). 
Furthermore, if the Higgs boson does not have the {\it exact} couplings to vector bosons as predicted by the SM, then the
 necessary cancelations will not occur and one will still observe an $E^2$ growth in the amplitudes until new physics 
 comes into play.  It is by measuring this growth that we can hope to observe the effects of physics beyond the SM, even
  in scenarios where we only see a light Higgs-like particle~\cite{Giudice:2007fh,Low:2009di}.

In this article we will introduce a novel technique designed to analyze VBF processes and observe the $E^2$ growth 
in longitudinal gauge-boson scattering amplitudes mentioned above.  We will begin by introducing our
notations and framework in Section~\ref{sec:setup}.
To motivate our new technique, in Section~\ref{sec:scaleuncert} we will update past analyses of VBF (specifically \cite{Bagger:1995mk} and~\cite{Butterworth:2002tt})
 taking into account the effects of parton showering and jet clustering.  We will show that these analyses, which infer the $E^2$
  amplitude growth from cross section increases, carry large $\mathcal{O}(100\%)$ uncertainties due to factorization-scale 
  ambiguities that affect jet tagging 
and vetoing.\footnote{While one might expect to eventually overcome these uncertainties through a combination of data-driven calibration 
and higher order calculations, the analysis we describe herein avoids them altogether.}  Then, in Section~\ref{sec:polar} we will 
 describe our technique designed to reduce 
these uncertainties by measuring the {\it relative} production of transverse to longitudinal modes, focusing on the fully reconstructable semi-leptonic 
decay of the $VV$ system.   We will demonstrate that this measurement is sensitive to anomalous Higgs-gauge couplings while at the same time 
being robust against the scale uncertainties that challenge cross section measurements.  In Section~\ref{sec:future} we will discuss 
potential improvements in the analysis and comment on other applications of the technique.  Section~\ref{sec:concl} contains our conclusions.

\section{Theoretical Setup}
\label{sec:setup}
In the formulation of a general effective theory of the SM-like Higgs sector \cite{Buchmuller:1985jz,Leung:1984ni,Barger:2003rs}
most of the operators are tightly constrained~\cite{Wudka:1994ny,Hagiwara:1992eh,Hagiwara:1993ck,Hagiwara:2002fs} 
because of their otherwise excessive contributions to the electroweak observables, such as the $\rho$-parameter,
oblique parameters, and triple gauge boson self-interactions. There are only two dimension-six operators that are genuine interactions 
in the Higgs sector not subject to the current experimental constraints, $\partial^\mu(H^\dagger H)\partial_\mu(H^\dagger H)$
and $(H^\dagger H)^3$.  We note that as both operators are composed from the singlet operator $H^\dagger H$ they may serve to probe
not only EWSB physics, but also other physics beyond the SM.
For a given theoretical framework, the coefficients of these operators
may be calculable~\cite{Giudice:2007fh}, and by measuring them we can hope to learn about any new physics.
  Even in some strongly coupled models for which these may not 
be calculable, the measurement of a non-zero value can give important clues to the structure of new physics.
Now, the second operator above does not have derivative couplings, so its effect on the behavior of the $V_L$ scattering
amplitudes at high energies should be sub-leading~\cite{Barger:2003rs}. We therefore  focus on the former and parameterize it (following~\cite {Giudice:2007fh}) as
\be
\mathcal{L}\supset\frac{c_H}{2f^2}\partial^\mu(H^\dagger H)\partial_\mu(H^\dagger H),
\ee
where the coefficient $c_H$ is naturally of $\mathcal{O}(1 - 4\pi)$ depending on whether the underlying theory is weakly 
or strongly coupled, 
and $f$ is the characteristic scale of new physics, typically expected to be round $4\pi v$ if the new physics is associated with EWSB.   

Upon expanding around the electroweak VEV $v$, this operator contributes terms which add to the kinetic terms of $H$. After imposing canonical 
normalization on the fields, the result is a modification to the Higgs couplings.  Ref.~\cite{Giudice:2007fh} parameterizes the 
resulting modified Higgs-gauge coupling in the zero-momentum limit as
\be
g_{\rm eff}=\frac{g^{}_{\rm SM}}{\sqrt{1+c_H\xi}}\approx g_{\rm SM}\left(1-\frac{c_H}{2}\xi\right)
\ee
where $\xi=v^2/f^2$.  This modified coupling has important phenomenological consequences because it deviates from the SM prediction. 
At high energies and for $\vert c_H\xi\vert \gtrsim 0.1$~\footnote{For smaller values of $\vert c_H\xi\vert$ the dominant non-SM effects 
enter as interference terms proportional to $c_H\xi$ rather than $(c_H\xi)^2$.  Also, in this case the anomalous energy dependence of the longitudinal cross section 
goes as $E^2$ instead of $E^4$.} this modification leads to an incomplete cancelation in  the amplitude 
for longitudinal vector boson scattering and the cross section grows as 
\be
\label{eq:llscat}
\sigma(V_L V_L\rightarrow V_LV_L)\approx\left(\frac{c_H}{2}\xi\right)^2\sigma(V_L V_L\rightarrow V_LV_L)_{\rm no-higgs}.
\ee 
which can be seen by considering the NGB scattering as shown in 
appendix~\ref{sec:scatamp}.
In what follows we will study means of measuring this behavior.  Note that, as discussed in 
appendix~\ref{sec:scatamp}, 
the $W^+_LW^-_L$ scattering amplitudes calculated in this framework violate perturbative unitarity when 
\be
s_{WW}\approx\frac{16 \pi v^2}{c_H\xi\left(1-{{c_H\xi }\over {4(1+c_H\xi)}}\right)}.
\ee
This is the point at which we expect new physics to come into play.  
In what follows we will limit our analyses to 
\be 
\sqrt{s_{VV}}<2~{\rm TeV}.
\ee
This corresponds to a coupling value $|c_H\xi | \sim 0.6$. We will take this as an upper limit for our analyses. 
Of course, looking beyond this energy range would be interesting and should be attempted at the LHC, 
but any deviation from the SM expectation  would no longer carry the same effective Lagrangian interpretation. 
Also, note that for larger couplings and lower scales of new physics some higher dimensional operators could become relevant
and it would be more appropriate to think of the $c_H\xi$ used in our analysis as parameterizing a new physics form factor, rather
than as the coefficient of a particular operator.

\section{Leptonic and Semi-Leptonic Channels Revisited: Scale Uncertainties}
\label{sec:scaleuncert}
The most straightforward way to probe the behavior of Eq.~(\ref{eq:llscat}) would be to measure the resulting 
increase in VBF cross section at higher energies.  This is a well studied topic, 
with many different analyses having
 been performed (see, for instance,~\cite{Bagger:1995mk,Butterworth:2002tt,Bagger:1992vu,Barger:1990py,
 Chanowitz:1998wi,Dicus:1990fz,Barger:1991ib,Hankele:2006ma,Bagger:1993zf,Gounaris:1993aj,Ballestrero:2009vw,Cheung:2008zh,He:2002qi,Zhang:2003it}).  
 Here we will revisit these analyses taking into account the effects of parton showering.  We
 will see that while the cuts developed in past analyses remain essential in suppressing background and isolating VBF signals, 
 one encounters large factorization scale uncertainties that affect rate measurements and must be
overcome to detect new physics in VBF.
%

Sophisticated acceptance  cuts have been developed to suppress the SM background and 
isolate the longitudinal gauge boson scattering in VBF processes at high energies.  It has been a common
practice to impose a high $p_T$ cut on the reconstructed gauge bosons
 or their decay products, require one or two forward (backward) energetic jets, and demand that the central detector 
 region remain relatively free of hadronic activity.  The first few cuts ensure that we observe hard scattering 
 processes with the gauge bosons emitted by energetic quarks~\cite{Cahn:1986zv,Barger:1988mr,Kleiss:1987cj}, 
while the last cut is designed to reduce background by taking advantage of the fact that VBF is a purely 
electroweak process with no color exchange~\cite{Barger:1994zq} and $V_LV_L$ scattering 
tends to produce fewer central jets than other electroweak processes \cite{Barger:1990py}.
Using this sort of cut, it was concluded \cite{Bagger:1995mk,Butterworth:2002tt} that reasonable sensitivity can be achieved for TeV scale 
strongly interacting new physics at the 14 TeV LHC with an integrated luminosity of 100 fb$^{-1}$.

We revisit  the $WW$  analyses with the theoretical framework as discussed in the previous section. We 
consider both fully leptonic \cite{Bagger:1995mk} 
and  semi-leptonic \cite{Butterworth:2002tt}  decays of the vector bosons. 
For the sake of illustration, we  concentrate on the $W^+W^-$ final state. 
Our VBF parton-level results are generated using the full $2\rightarrow 6$ matrix element for
\be
q q' \to q q' W^+W^- \to  q q'\  \ell^\pm  \nu\  f \bar f' ,
\ee
without making the effective $W$ approximation~\cite{Cahn:1983ip,Kane:1984bb,Dawson:1984gx}. In so doing, 
wherever appropriate, we have included other $\mathcal{O}(\alpha_{EW}^6)$ processes as background to the channels. 
Our PDFs are those of MRST2004~\cite{Martin:2004ir}.
To generate the jet-level samples we shower parton-level results using \texttt{Pythia 6.4.21}~\cite{Sjostrand:2006za} 
with a virtuality ordered shower, cluster the visible final state particles into $0.1\times0.1$ $y-\phi$ cells between $-5\leq y\leq 5$, 
and produce $R=0.7$ anti-$k_T$~\cite{Cacciari:2008gp} jets using FastJet~\cite{Cacciari:Fastjet}.  To sample PDFs and shower 
our results we must choose a factorization scale for the gauge boson 
scattering processes.  The natural choice of the factorization scale is of the order of $m^{}_W$, 
with corrections from the $p_T$ of the scattering quarks.
We parameterize the choice of scales via
\be
\mu^2=\beta^2\left(m_W^2+\frac{1}{2}\sum_{{\rm jets}}p_T^2\right),
\label{eq:scale}
\ee
where $\beta$ is an $\mathcal{O}(1)$ parameter. 

We begin by adopting the selection cuts of~\cite{Bagger:1995mk} to study the fully leptonic $W^+W^-$ final state, as
detailed in Table~\ref{tab:tao_cuts}. 
Using these cuts, we calculate the parton level cross sections for a light Higgs 
scenario\footnote{Here and henceforth, we will take a light Higgs boson mass as  $m_H=100~{\rm GeV}$ for illustration and for
comparing with the early studies in the literature. This will make no numerical difference with other $m_H$ values as long as it is well below
the $2 m^{}_W$ threshold.} with various anomalous couplings parameterized by $c_H^{} \xi$. 
The parton level results for a few representative scale choices 
of $\beta$ are listed on the  left-hand side of Table~\ref{tab:taocutcs}. They are consistent with those of~\cite{Bagger:1995mk}.  
At this level, the uncertainty in rate is only around $\mathcal{O}(10\%)$, which
can be attributed entirely to the PDFs.
When we include showering, hadronization and jet clustering, the scale $\mu$ dictates the virtuality 
at which the parton shower is started, in addition to controlling the sampling of PDFs.  On the right-hand side in Table~\ref{tab:taocutcs}, 
we present the cross sections for the showered and clustered results  with a few representative scale choices.
We see that the uncertainties can now be as much as $\mathcal{O}(100\%)$.  
This is because small changes in $\mu$ result in large changes in the behavior of the associated forward jets.  A 
higher value of $\mu$ could lead to harder radiation that will sink forward jets below the tagging criteria, or it could 
lead to the parton-shower emission of a veto jet. 
As the uncertainties from varying the 
scale ($\beta=0.5 - 2.0$) would normally set the systematic theoretical errors,
%
such large uncertainties in rate would make it difficult to distinguish 
the presence of anomalous couplings, even for large values of $c_H\xi$. 

\TABLE[t]{
\parbox{\textwidth}{
\begin{center}
\begin{tabular}{|c|c|}
\hline
Leptonic Cuts & Jet Cuts\\
\hline
$|y(l) |< 2.0$ & $E(j_{\rm tag}) > 0.8~{\rm TeV}$\\
$p_T(l) > 100~{\rm GeV}$ & $3.0 < |j_{\rm tag}| < 5.0$ \\
$\Delta p_T(ll) > 440~{\rm GeV}$ & $p_T(j_{\rm tag}) > 40~{\rm GeV}$\\
$\cos\phi_{ll} < -0.8$ & $p_T(j_{\rm veto}) > 30~{\rm GeV}$\\
$M(ll)> 250~{\rm GeV}$ & $ |y(j_{\rm veto})| < 3.0$\\
 \hline
\end{tabular}
\end{center}
\caption{\label{tab:tao_cuts} The cuts of~\cite{Bagger:1995mk} for the leptonically decaying $W^+W^-$ final state.  
The signal selection requires that we tag at least one  jet ($j_{\rm tag}$) and to veto extra central jets ($j_{\rm veto}$).}
}}
\TABLE[t]{
\parbox{\textwidth}{
\begin{center}
\begin{tabular}{|c|c|c|c|c|c|c|c|c|}
\hline
&\multicolumn{3}{c|}{Parton Level [fb]} &\multicolumn{3}{|c|}  {Jet Level [fb]} \\
\hline
${c_H} \xi $& $\beta=0.5$ & $\beta=1.0$& $\beta=2.0$& $\beta=0.5$ & $\beta=1.0$& $\beta=2.0$\\
\hline
0.4&  0.015 & 0.013& 0.012 & 0.015 & 0.009 & 0.005  \\
0.2 & 0.013 & 0.011 & 0.010  & 0.013 & 0.006 & 0.004 \\
0.0 & 0.011& 0.090& 0.008 &0.012 & 0.007& 0.004 \\
 \hline
\end{tabular}
\end{center}
\caption{Cross sections [fb] for VBF with $W^+W^-$ final states decaying into $e$ and $\mu$ for various anomalous 
Higgs-gauge couplings and at different factorization scales parameterized by Eq.~(\ref{eq:scale}).  
The cuts used to generate these results are those of~\cite{Bagger:1995mk}.  
The set of cross sections on the left are computed at parton level,  while those on the right correspond to results 
after the parton shower  and hadronization. 
 \label{tab:taocutcs}}
}}

%
We next explore the situation for the semi-leptonic mode of  $W^+W^-$ decay. 
We employ cuts inspired by~\cite{Butterworth:2002tt} as shown  in Table~\ref{tab:butt_cuts}.  
The results of this analysis are shown in Table~\ref{tab:buttcutcs}, again demonstrating a relatively stable signal
at the parton level (left-hand panels) and an $\mathcal{O}(100\%)$ uncertainty at the jet level (right-hand panels). 
As with the fully leptonic system considered above, the large uncertainty is once again attributable to the parton-shower treatment 
of the forward jets using different scales.
\TABLE[t]{
\parbox{\textwidth}{
\begin{center}
\begin{tabular}{|c|c|}
\hline
Pass conditions& Veto conditions\\
\hline
$E(j_{{\rm tag}}) > 300~{\rm GeV}$&$p_T(j_{\rm mini}) > 25~{\rm GeV}$ \\
$2<|y(j_{{\rm tag}})| <5$&$|y(j_{\rm mini})| < 2$\\
$p_T(j_{\rm tag})>20~{\rm GeV}$&$130~{\rm GeV} < m_{WJ} < 240~{\rm GeV}$ \\
$p_T(W_{\rm recon.})>320~{\rm GeV}$& \\
$|y(W_{\rm had})| < 4$ & \\
 \hline
\end{tabular}
\end{center}
\caption{\label{tab:butt_cuts} $W^+W^-$ semi-leptonic decay cuts inspired by~\cite {Butterworth:2002tt}.  
These require two tagged jets ($j_{\rm tag}$) and two reconstructed $W$s ($W_{\rm recon.}$).  If the events 
contain two soft, central jets ($j_{\rm mini}$) they are vetoed.  The cut on the jet-$W$ invariant mass is 
designed to reduce top quark backgrounds.}}}
\TABLE[t]{
\parbox{\textwidth}{
\begin{center}
\begin{tabular}{|c|c|c|c|c|c|c|c|c|}
\hline
&\multicolumn{3}{c|}{Parton Level [fb]} &\multicolumn{3}{c|} {Jet Level [fb]} \\
\hline
${c_H} \xi $& $\beta=0.5$ & $\beta=1.0$& $\beta=2.0$& $\beta=0.5$ & $\beta=1.0$& $\beta=2.0$\\
\hline
0.4  &  0.95 & 0.81 & 0.73 & 0.53 & 0.38 & 0.26\\
0.2  & 0.82 & 0.72 & 0.64  & 0.43 & 0.33 & 0.24 \\
0.0  & 0.73 & 0.64 & 0.57 &0.40 & 0.29& 0.21 \\
 \hline
\end{tabular}
\end{center}
\caption{Cross sections [fb] for VBF with the semi-leptonic decay of $W^+W^-$ for various anomalous 
Higgs-gauge couplings and at different factorization scales parameterized by Eq.~(\ref{eq:scale}).    
The cuts used to generate these results are those of Table~\ref {tab:butt_cuts}. \label{tab:buttcutcs}}
}}

We find that VBF cross sections with complicated kinematical cuts 
are extremely sensitive to the exact scale choice 
one uses, and it is insufficient to simply characterize the hard scattering with a rough estimate of $\mu$.   
While we have only considered the effects of varying $\mu$ on signal rates, the background is also 
susceptible to these uncertainties.  Even without considering the effects of the veto cut, which can only
increase sensitivity to $\mu$,  
it's natural to associate $\mathcal{O}(2\times)$ $K$-factors with high multiplicity QCD events.
%
Therefore, before drawing any conclusion about the presence of new physics one would have to understand these systematics.
In principle, the theoretical uncertainty may be reduced through higher order calculations that can give us a 
better idea of the appropriate scale choice. Substantial efforts would be needed both in theory and in experiments
before to bring this uncertainty under control.
%
With this in mind, in the next section we will present a new tool to circumvent the difficult
issue of the factorization-scale dependence.  
\section{Polarization Measurements}
\label{sec:polar}
With the uncertainties detailed above as our motivation, we propose  a new technique 
to probe the anomalous couplings in a robust way.  Our basic idea is to look for the {\it relative} 
increase in longitudinal vector boson production by comparing it to the production of transverse 
modes.  Unlike the overall cross section, which is sensitive to the behavior of the forward jets, the 
relative transverse to longitudinal production rates should be stable against different scale choices 
because it depends only on the $VV\rightarrow VV$ scattering amplitude.  
\FIGURE[t]{
\includegraphics[scale=0.9]{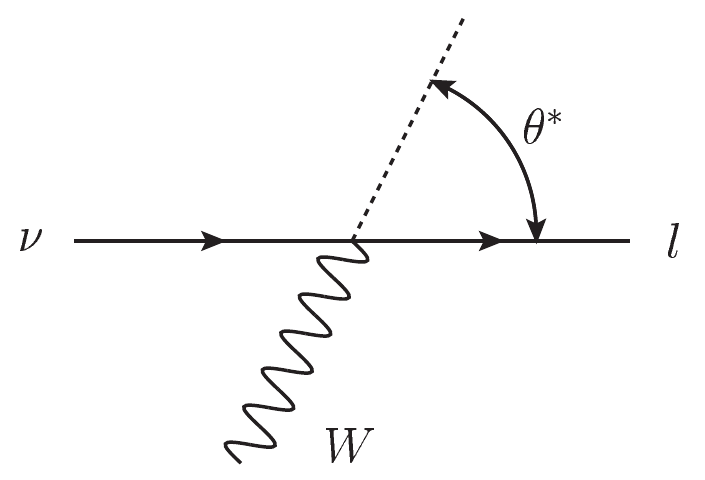} 
\caption{
\label{fig:pol_axis} The polarization axis used to measure $\theta ^\ast$.  Note that this is 
measured in the rest frame of the $W$, and the $W$ direction of motion is defined with respect to the $WW$ center of mass.}  }
To measure the polarization of a vector boson we need to reconstruct the four-momenta of its 
decay products and measure their distribution with respect to a polarization axis.  If one chooses 
the polarization axis to be the gauge boson direction of motion (Fig.~\ref{fig:pol_axis}), then a simple 
spin-analysis predicts that in the $V$ rest frame the transverse and longitudinal polarizations will be distributed as
\be
\label{eq:pol_fcns}
P_\pm(\cos \theta ^\ast)=\frac{3}{8}(1\pm\cos \theta ^\ast)^2,\ P_L(\cos \theta ^\ast)=\frac{3}{4}(1-\cos^2\theta^\ast)
\ee
where $\theta^\ast$ denotes the angle between the parton and the gauge boson direction of motion in the 
gauge boson rest frame.  

To measure these distributions experimentally, we need to fully reconstruct the gauge boson pair 
center of mass and each gauge boson's direction of motion in this frame.  To accomplish this we will 
focus on the semi-leptonic decay channel of the $VV$ system as this allows full reconstruction 
of the system while minimizing the SM background by requiring leptons and missing 
energy. The semi-leptonic channel also significantly increases the signal event rate.
 For this we will rely upon jet substructure techniques to reconstruct the hadronically decaying 
gauge boson~\cite{Butterworth:2002tt}.  We will focus on studying the $W^+W^-$ final state, although we will 
take into account the background from other VBF processes like $W^\pm W^\pm$ and $W^\pm Z$ that enter 
because we can not distinguish the sign of a hadronically decaying vector, nor can we 
always distinguish a hadronically decaying $W$ from a $Z$.  Later in this section 
we will comment on the SM $\mathcal{O}(\alpha_S^2)$ and $\mathcal{O}(\alpha_S^4)$ backgrounds.

\FIGURE[t]{
\includegraphics[scale=0.35]{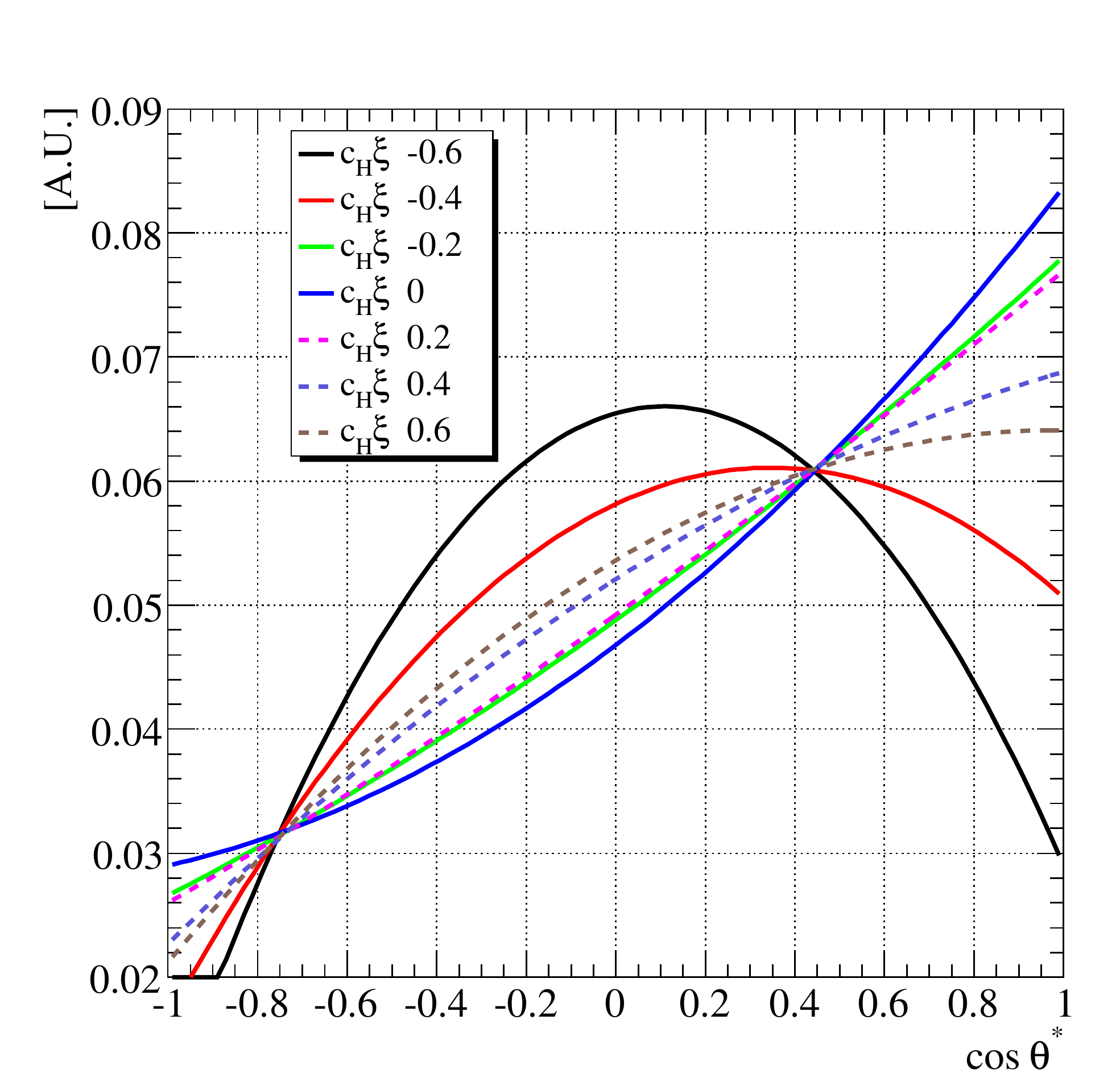} 
\includegraphics[scale=0.35]{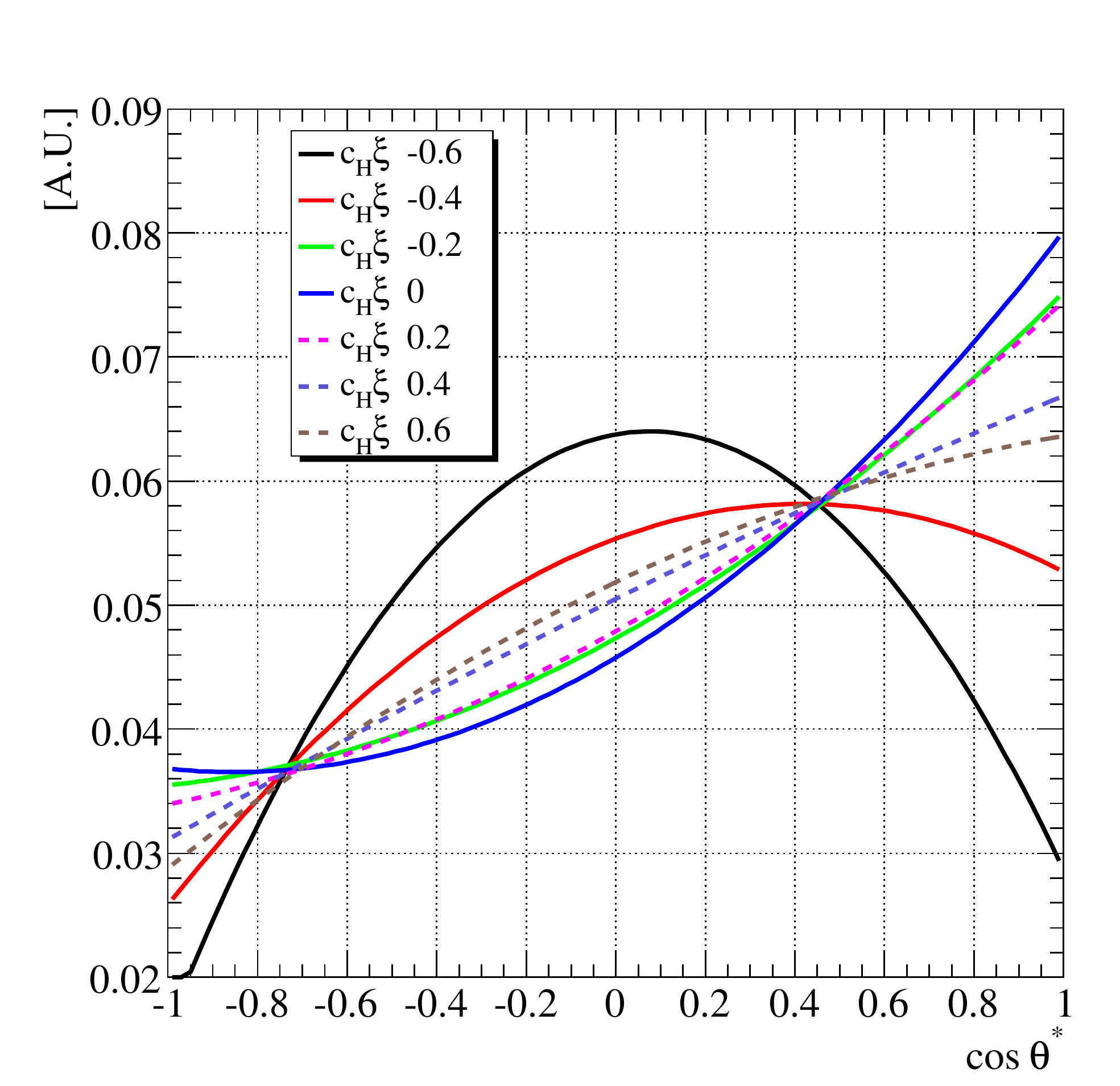} 
\caption{
\label{fig:ww_lep}The distributions of $\cos \theta ^\ast$ for different anomalous couplings at parton level (left) and for fully showered, 
hadronized, clustered, and reconstructed events (right). All distributions are normalized to the same area.}}

\subsection{Leptonic Polarization}
We begin with the polarization analysis for the leptonic side of the decay.  We first study the 
parton-level results, then we will turn on the full simulation (parton-showering and jet clustering) to 
see that they are largely unchanged.  The cuts we use are those in Table~\ref{tab:butt_cuts}. 

Before proceeding further, we encounter a subtlety in the reconstruction of the leptonic system: 
While the neutrino four-momentum is constrained by the on-shell $W$ condition, it is only 
determined up to a discrete ambiguity. One finds two candidate four-momenta at the same 
azimuthal angle but separated from the charged lepton by a fixed rapidity difference.  In what 
follows we will simply use the average $\cos\theta^\ast$ value from both solutions as an approximation 
of the true value.  This is acceptable because we are working in a boosted regime where the difference 
in rapidity between neutrino and lepton is small, making the curvature effects from the $(y,\phi)$ system 
sub-leading.
The resultant distributions are shown in Fig.~\ref{fig:ww_lep}, at parton level (left panel) and after 
the hadronization (right panel).  The characteristic shapes with different couplings  are quite distinctive.
In Table~\ref{tab:lep_pol_frac} 
we compute the cross section for each anomalous coupling and fit it to the transverse and longitudinal 
distributions of Eq.~(\ref{eq:pol_fcns}) using
\be
P(\cos\theta^\ast)=f_L P_L(\cos\theta^\ast)+f_+P_+(\cos\theta^\ast)+f_-P_-(\cos\theta^\ast)
\ee
where the $P$ are normalized probability distributions of $\cos\theta^\ast$ and the $f$ are subject to
 the constraint $\sum f=1$.  As one can see from comparing the jet and parton level figures, the results 
 are remarkably stable under a full simulation.
\TABLE[t]{
\parbox{\textwidth}{
\begin{center}
\begin{tabular}{|c|c|c|c|c|c|c|c|c|}
\hline
&\multicolumn{2}{c}{Leptonic $W$} &\multicolumn{2}{|c|}{Hadronic $W$} &\\
\hline
${c_H} \xi $&$f_L^{P}$& $f_L^J$&$f_L^{P}$& $f_L^J$& $\sigma$ [fb]\\
\hline
-0.6& 0.77& 0.74 & 0.71 & 0.55& 3.38 \\
-0.4 & 0.58 & 0.52& 0.49 & 0.40 & 1.12\\
-0.2 & 0.33 & 0.30& 0.23 & 0.24 & 0.60\\
0.0 & 0.27 & 0.25 & 0.17 & 0.22 & 0.62\\
0.2 & 0.34 & 0.31 & 0.24 & 0.26& 0.65\\
0.4 & 0.42 & 0.39& 0.32 & 0.32  & 0.73\\
0.6 & 0.46 & 0.42 & 0.40 & 0.38 &0.87\\
 \hline
\end{tabular}
\end{center}
\caption{The fraction of longitudinally polarized vector bosons for different 
anomalous couplings at parton level $f_L^{P}$ and jet level $f_L^{J}$, reconstructed in hadronic and leptonic decays.  Also listed are 
the jet-level cross sections\label{tab:lep_pol_frac}\label{tab:hadlongfrac}. These results are after the cuts of Table~\ref{tab:butt_cuts}.}
}
}

In Figure~\ref{fig:lep_reach} we plot the projected event distributions and associated statistical errors both for the 
SM and for an anomalous scenario with $c_H\xi=-0.4$, given $100\ {\rm fb}^{-1}$ of luminosity.  
The shape difference between the two samples is clearly visible.  To estimate the luminosity necessary to 
probe a given coupling, one can use that the signal scales roughly as $(c_H\xi)^2$, as discussed before.
However, the precise reach of the LHC in discerning anomalous couplings 
will require a more thorough accounting of background (we discuss this further in Section~\ref{sec:bg}). Further, 
we have not made an effort to optimize the statistical power of the analysis and there are
other channels that contribute to the signal, such as $W^+W^+, W^\pm Z$ and $ZZ$.  In addition, one can 
extract more information from each event, as we will now see.

\FIGURE{
\includegraphics[scale=0.5]{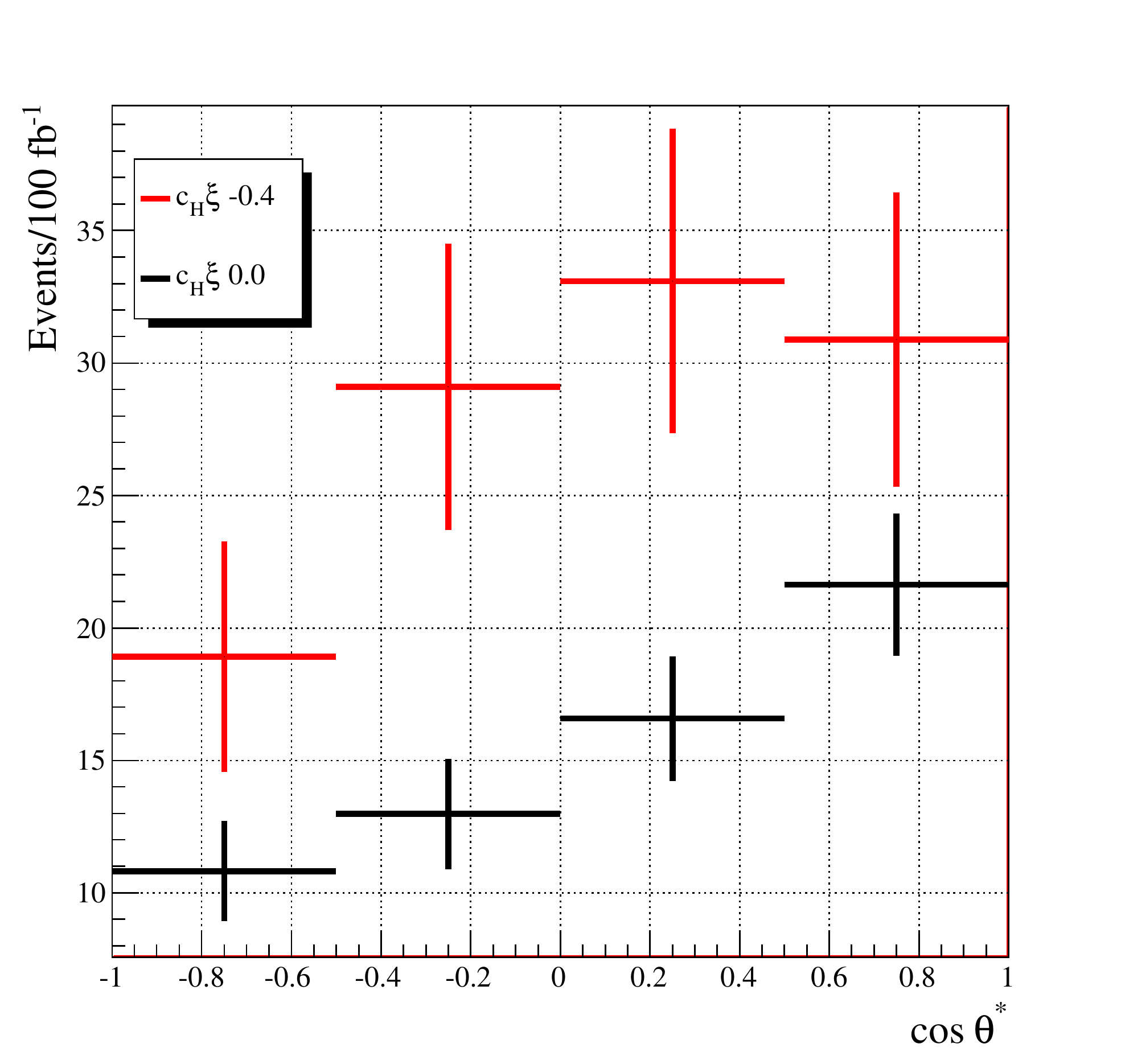} 
\caption{Projected distribution and associated statistical uncertainties of $\cos\theta^\ast$ for the leptonically decaying vector using $100\ {\rm fb}^{-1}$ of luminosity.
\label{fig:lep_reach}}
}

\subsection{Hadronic Polarization}
It is possible to further improve the discriminating power of polarization by considering both sides of the $VV$ system together;
by looking for the expected {\it correlation} between both states one can hope to gain additional discriminating power.  

\FIGURE[t]{
\includegraphics[scale=.35]{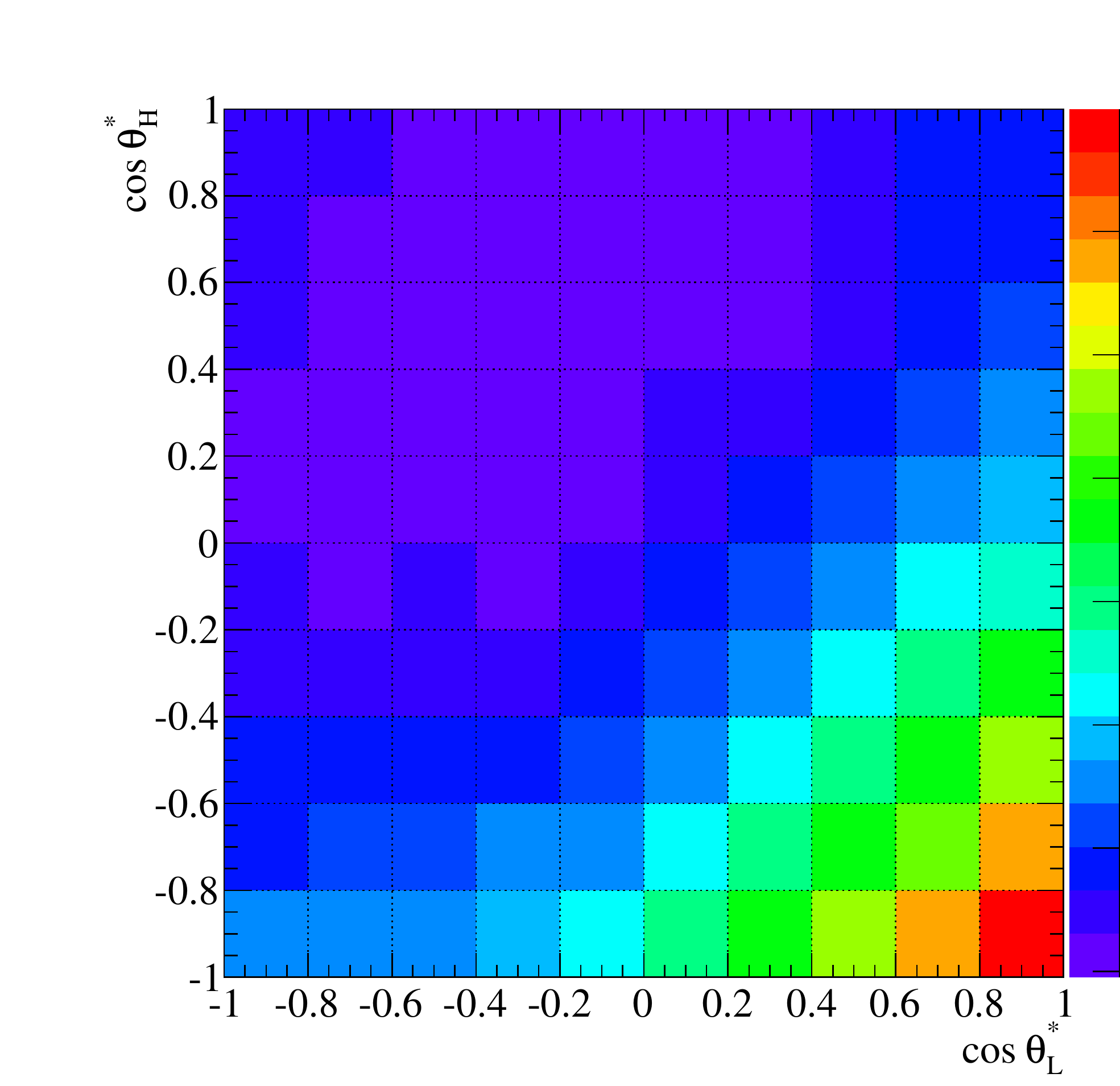} 
\includegraphics[scale=.35]{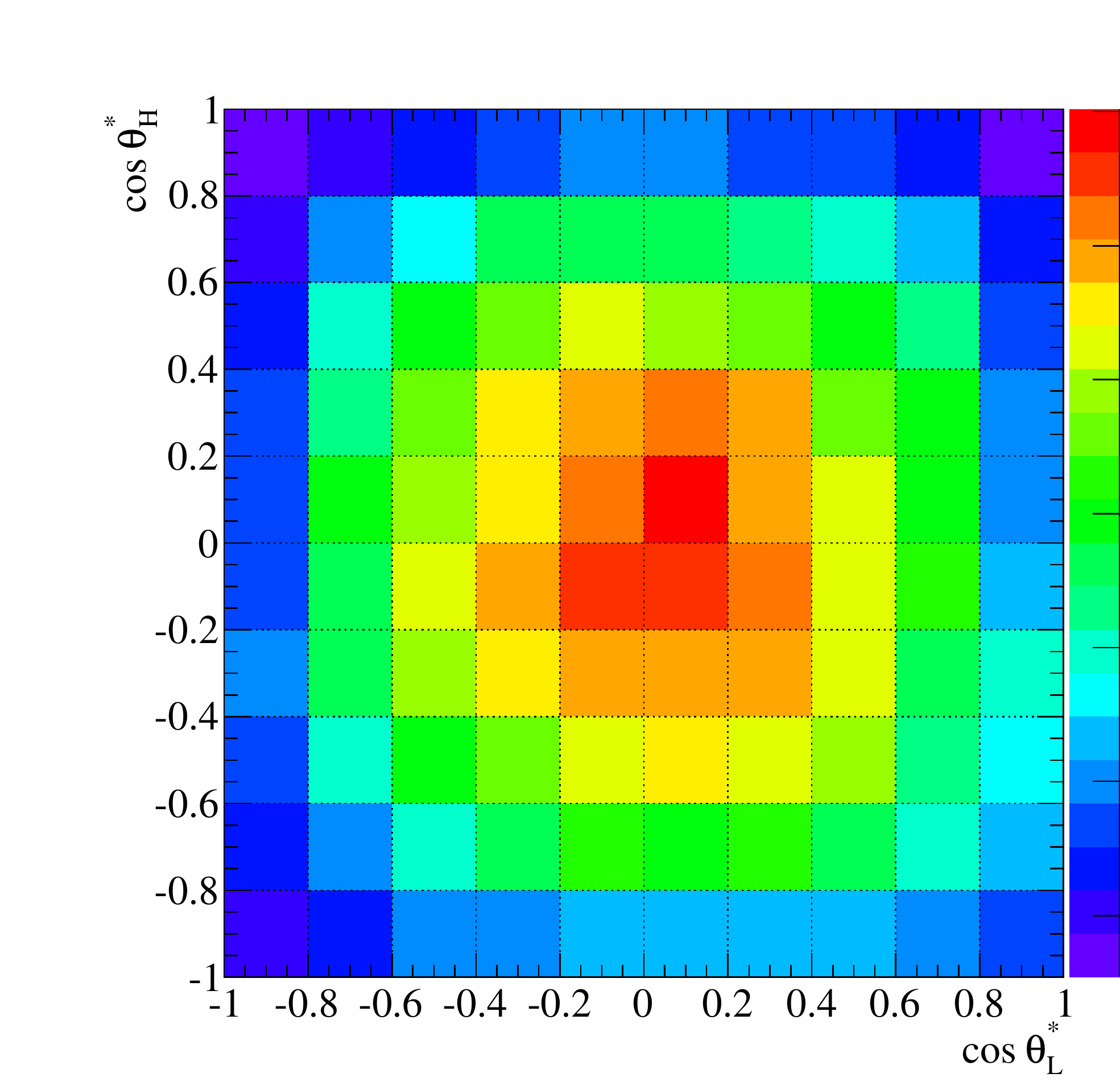} 
\caption{\label{fig:2d_lep} Distributions of $\cos \theta ^\ast$ at parton-level for both sides of the $VV$ system 
(labeled with subscript $H$ and $L$ for hadronic and leptonic decays, respectively).  
The plot on the left is for the Higgs with SM couplings, while the one on the right is for $c_H\xi=-0.6$.  The scale 
is individually normalized for each plot, going from violet to red as the concentration of events increases.  The scaling of the color gradient 
on the right side of each plot is linear.}}

To see the correlation effect, consider Fig.~\ref{fig:2d_lep}, which shows the parton-level $\cos \theta ^\ast$ 
distributions for both sides of the $VV$ system in SM and non-SM scenarios. For now, we plot  $\cos\theta^\ast$ 
on the hadronic side for the down-type quarks.
In the non-SM scenario we see a rapid rise in the central region of the plot near $\cos \theta ^\ast\approx0$.  This 
indicates that the results are correlated; when we see a $V_L$ it is likely to be accompanied by a $V_L$ because
only the $V_LV_L$ final state sees the $E^2$ growth characteristic of with non-SM effects.
In practice the situation is slightly more complicated because we cannot label the light quark states once they 
shower and hadronize (e.g. we cannot distinguish a $u$ from a $d$), so the distributions we measure are symmetrized.  
However, the distributions still carry additional discriminating power, as one can see from the distributions in 
Fig.~\ref{fig:had_decay} and Fig.~\ref{fig:2dhadsymm}, and Table~\ref{tab:hadlongfrac}.  Note that in fitting the 
symmetrized distributions we only fit to data from $0 < |\cos\theta^\ast| < 0.7$.  In the regime where  
$|\cos\theta^\ast| \gtrsim 0.7$ one subject becomes very soft and the technique breaks down (although, 
of course, the leptonic analysis still works here).

\FIGURE[t]{
\includegraphics[scale=.35]{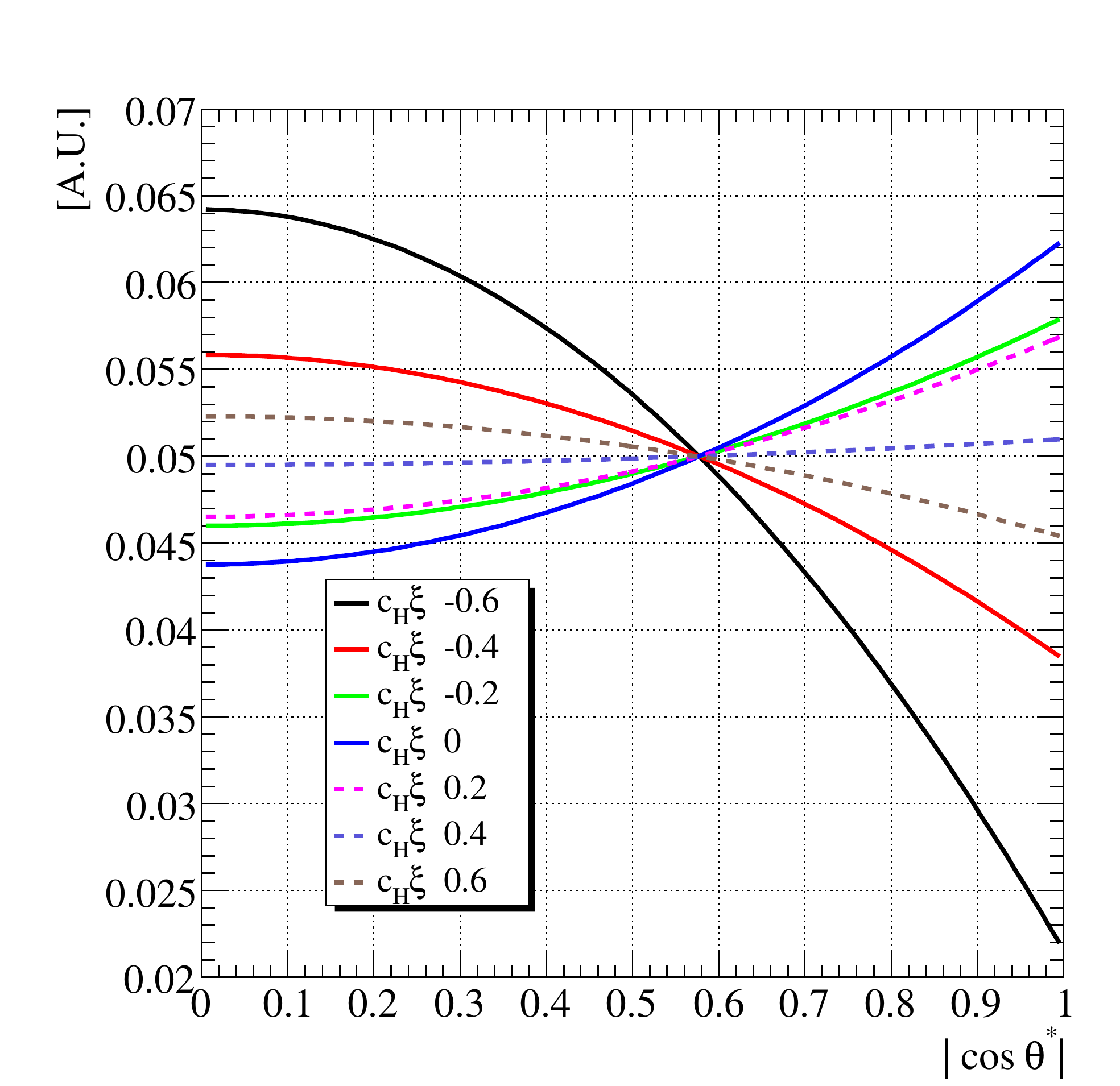} 
\includegraphics[scale=.35]{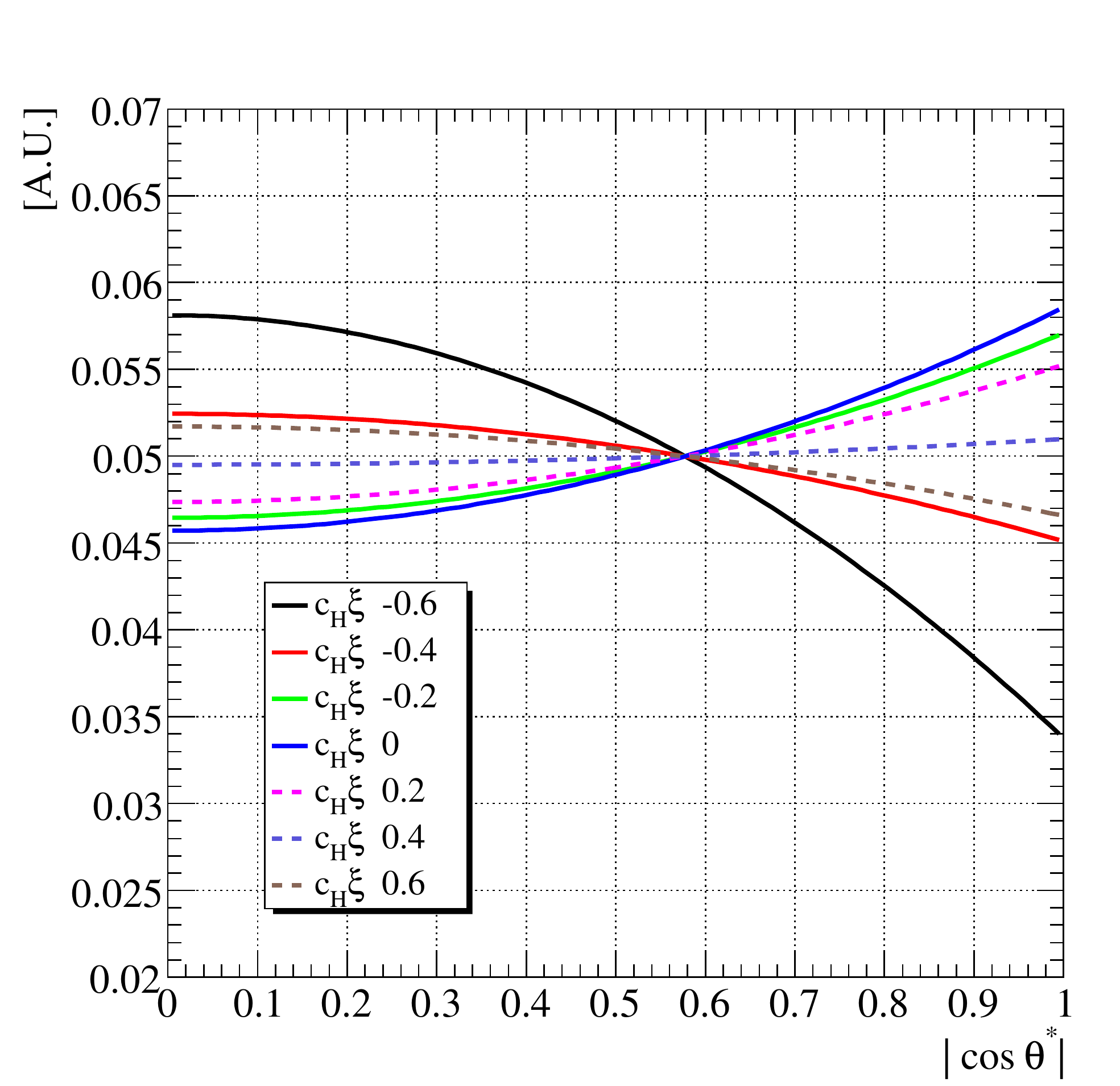} 
\label{fig:had_decay} 
\caption{Distribution of $|\cos \theta ^\ast |$ at different anomalous couplings for hadronically decaying $W$s using parton level samples (left) and 
fully showered, hadronized samples (right).  Note that the distributions differ more at high values of $\cos \theta ^\ast$ 
because this is the region in which one jet is relatively soft.}
}

To perform this analysis  we had to look at the hadronically decaying $V$ using subjet techniques (for a 
short overview of jet algorithms and their behavior, see appendix~\ref{sec:jetreview}).  
In particular, we used the $k_T$ algorithm~\cite{Catani:1993hr,Ellis:1993tq} with $R=0.25$ to cluster the 
constituents of each hadronically decaying gauge boson, using the two most energetic subjets (as measured in the $VV$ 
center of mass frame) for our analysis.  Note that rather than identifying our subjets through a 
C/A~\cite{Dokshitzer:1997in,Wobisch:1998wt} or $k_T$-like unwinding~\cite{Butterworth:2008iy,Thaler:2008ju}, 
we used fixed small cones (i.e. small $R$).  Otherwise, the subjets encompass a large area and become more susceptable
to contamination from initial state radiation, multiple interactions, and event pileup.  The choice
of a small cone seems to result in a better reconstruction of events, especially at high values of 
$\cos\theta^\ast$ when there is a large difference in the subjet $p_T$s.  Furthermore, we use $k_T$  rather than 
anti-$k_T$ to form our subjets because it more accurately reconstructs the softer jet in situations where the jets are nearly collinear (see appendix~\ref{sec:jetreview}).  

One important thing to consider in the subjet analysis is that the results are not as robust in going from 
matrix-element to parton shower as were the leptonic results; the curves change shape (compare the parton and jet level results 
for both sides of the decay in Table~\ref{tab:lep_pol_frac}).  This is 
because the diffuse nature of the subjets makes them difficult to resolve when they become collinear 
and/or soft.  We note, however, that at the LHC we can expect to calibrate subjet measurements for 
boosted hadronic $W$s with large SM samples, and while the parton-level to jet-level results may vary, 
the correspondence should eventually be well understood.  Thus the leptonic gauge boson analysis is likely 
to be the first tool used, but the hadronic analysis can be added later on. 
\FIGURE[t]{
\includegraphics[scale=.35]{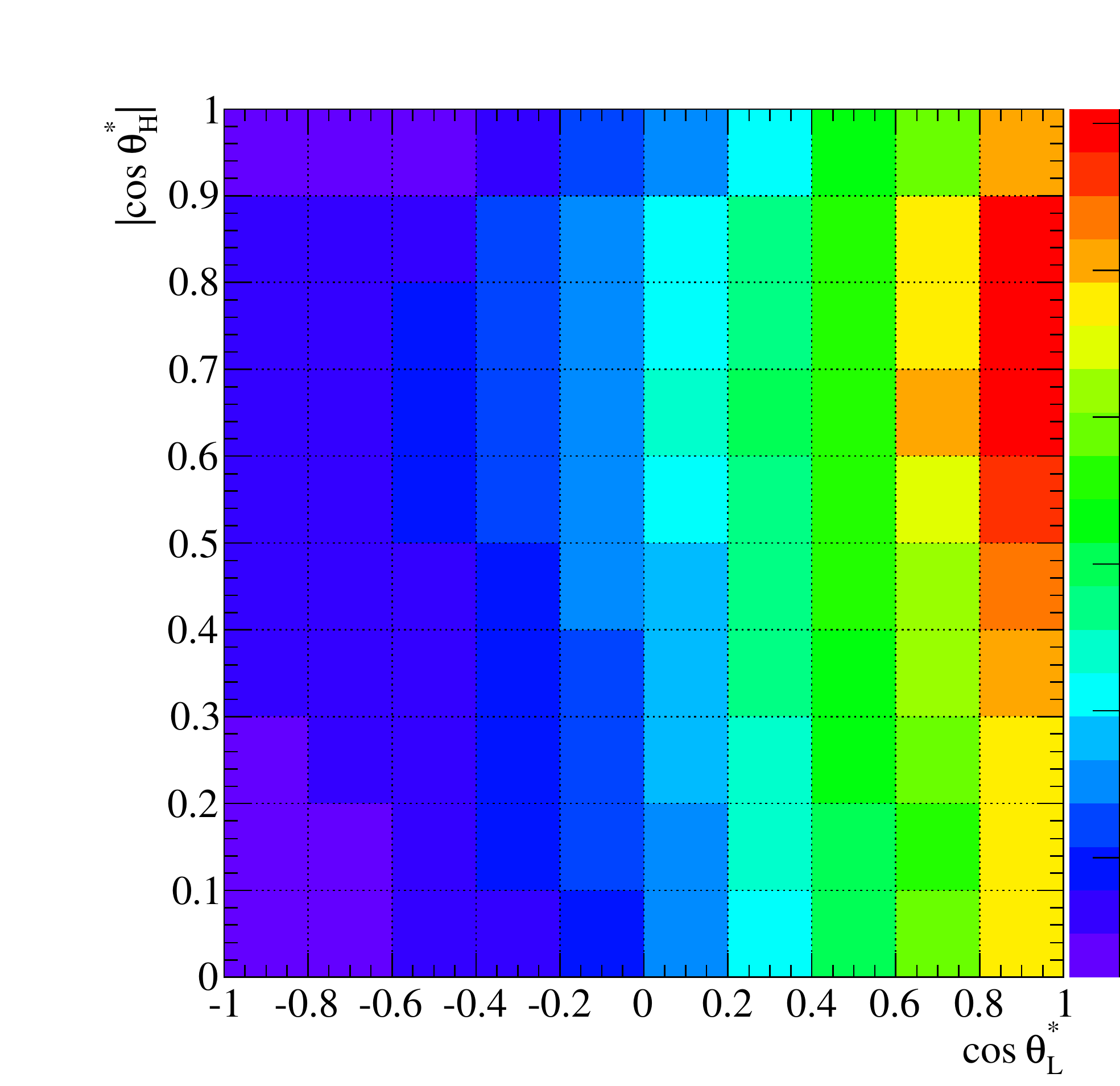} 
\includegraphics[scale=.35]{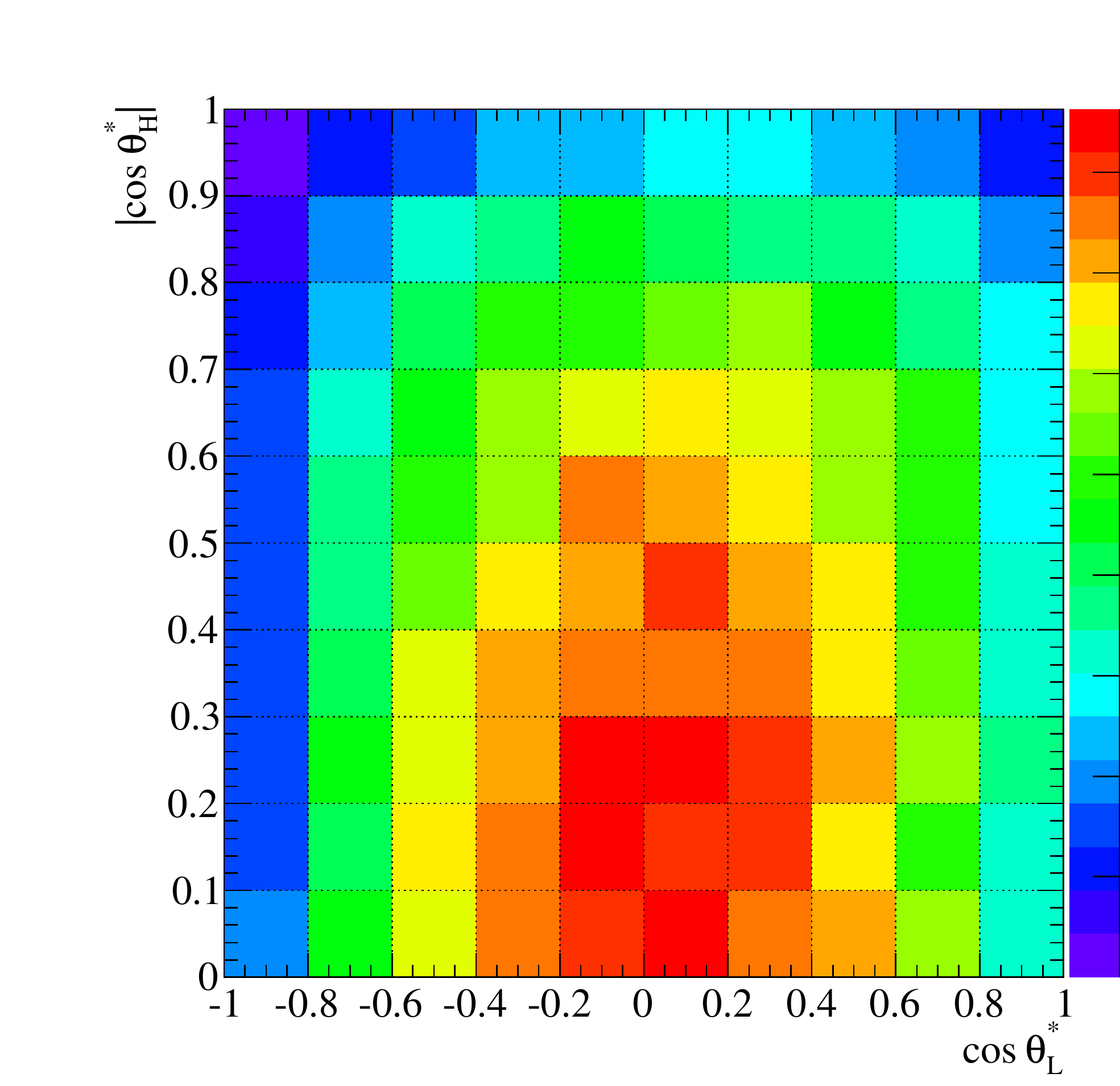} 
\label{fig:2d_had_decay} 
\caption{Jet-level distribution of $\cos \theta ^\ast$ (labeled with subscript $H$ and $L$ for hadronic and leptonic decays, respectively)  
for the SM Higgs (left) and Higgs with $c_H\xi=-0.6$ (right).  
The scale is individually normalized for each plot, going from violet to red as the concentration of events increases. The scaling of the color gradient 
on the right side of each plot is linear.\label{fig:2dhadsymm}}
}

\subsection{Scale Uncertainty}
We now explore the sensitivity of our results to the choice of the factorization scales.
As before, we vary the scale choice as parameterized by Eq.~(\ref{eq:scale}). 
We present  our fully showered results for the leptonically decaying $W$ in Table~\ref{tab:long_frac_scale_dep} using different scale choices.  
One can see from 
these results that the relative longitudinal to transverse fraction is a robust measure of anomalous 
couplings, steady across different values of $\beta$.  The reason for this is clear: The showering and hadronization
processes of the forward jets are independent of the behavior of the polarizations and decays of the gauge bosons.  
Whatever happens with the forward jet-tagging and the central jet-vetoing 
is isolated from the polarization measurements of the final decay products, hence the robust behavior.
\TABLE[t]{
\parbox{\textwidth}{
\begin{center}
\begin{tabular}{|c|c|c|c|c|}
\hline
&\multicolumn{3}{c|}{Longitudinal Fraction} \\
\hline
${c_H} \xi $&$\beta=0.5$& $\beta=1.0$ & $\beta=2.0$ \\
\hline
0.0 & 0.25 & 0.26 & 0.25\\
0.2 & 0.33 & 0.33 & 0.33 \\
0.4 & 0.40 & 0.40 & 0.41\\
 \hline
\end{tabular}
\end{center}
\caption{The reconstructed fraction of longitudinally polarized leptonically decaying vector bosons measured 
for various anomalous couplings and at different factorization scales.
\label{tab:long_frac_scale_dep} }
}
}
\subsection{Background}
\label{sec:bg}
In our matrix element calculations, we have included the irreducible electroweak $VVjj$ backgrounds. 
A full treatment of the other background from QCD and top quarks 
is beyond the scope of this work, but here we give some rough estimates 
and discuss how background processes might have an effect upon the calculations.
%
With the sophisticated acceptance cuts as adopted here, 
estimates of the background generally place it at or below the signal level. 
For the semi-leptonic modes of our interest in~\cite{Butterworth:2002tt}, 
whose cuts we mimic, the background is calculated to be roughly one quarter of the signal rate, so we 
expect the enhanced signal distributions to be discernible.  
%
We would like to emphasize that the robust feature of the polarization measurements should still persist when
comparing with the background events, in which the leptonic decaying $W$ is mostly transversely polarized,
and the di-jets from QCD have no particular angular preference.
%
%
Thus further separation of the backgrounds may be possible by extending this technique to the two-dimensional angular
plots where one looks for correlations as shown in the last section. 

\section{Future Directions}
\label{sec:future}
Here we discuss some further improvements one could make in reconstructing the hadronically 
decaying gauge bosons, along with some additional applications and extensions of our technique.

\subsection{Improving Hadronic Reconstruction}
Comparing the columns of Table~\ref{tab:lep_pol_frac}, one can see that the polarization 
of the hadronically decaying gauge boson was reconstructed less accurately than the gauge boson that decayed leptonically.  
This is partly because at higher boosts the gauge boson decay products become more collinear and are consequently 
more difficult to resolve individually.  While it is tempting to simply exclude this region from the analysis, the 
highly boosted region is precisely where we expect to see the greatest effects of new physics, so it is 
worthwhile to try to improve reconstruction in this regime. 

One approach for better resolving the subjets would involve letting the subjet radius scale as $1/p_T$, so that 
the size of the subjets is adjusted to naturally account for their kinematics. This sort of scaling was explored 
by~\cite{Krohn:2009zg}, although in a non-subjet context.  We note that it is likely one would have to adapt 
the $1/p_T$ behavior to account for the finite calorimeter segmentation and crowded subjet environment if this approach were adopted.

Another approach, along a different line, would involve abandoning subjet construction at very high boosts 
in favor of an energy splitting technique.  Such a technique would split the jet from a hadronically decaying 
gauge boson into two pieces and consider the degree of energy sharing between them using something akin to 
the $z$-variable of~\cite{Thaler:2008ju}:
\be
z=\min(E_A,E_B)/E_{\rm tot}.
\ee
For $\cos\theta^\ast\approx0$, where both daughters are transverse to the $V$ direction of motion we would 
expect $z\approx0.5$, while for $\cos\theta^\ast\approx1$ we expect $z\approx 0$.  We avoided this technique 
earlier in our polarization analysis because it is not boost invariant, and an alternative definition of $z$ using 
$p_T$ instead of $E$ would still be sensitive to the orientation of the $V$ decay.  However, these effects go 
as $m_W/p_T$, and so are marginalized as we consider higher boosts.  Therefore, it might become advantageous 
to transition to a $z$ variable in the highest $p_T$ ranges.

\subsection{Additional Applications}
Throughout this article we have focused on scenarios with a light, SM-like Higgs because they can be very difficult 
to distinguish from the SM, making VBF analyses particularly useful.  However, the VBF polarization measurements 
we have discussed could be of use in studying other models of EWSB.  

Consider, for example, the class of models where $V_LV_L$ scatter amplitudes are unitarized by the exchange of 
new spin-1 states (see Fig.~\ref{fig:spin_one_unit}). Models in this class include the so-called Higgsless models, 
where  higher dimensional gauge bosons unitarize the longitudinal amplitudes, and technicolor, which employs 
the technirho for this 
purpose~\cite{Csaki:2003zu,Csaki:2003dt,Chivukula:2003kq,Chivukula:2006cg,Nomura:2003du,Susskind:1978ms,Weinberg:1979bn}.  

Phenomenological studies of these models~\cite{Belyaev:2008yj,Birkedal:2004au} often study the VBF production 
of the new resonances because the heavy spin-1 states involved in restoring unitarity must couple strongly to the 
longitudinal modes of the electroweak gauge bosons.  While it would be striking to see a bump in the $VV$ invariant 
mass to reconstruct a resonance, seeing this correlated with a measured increase in the longitudinal production would give a
 true smoking gun signal.  Furthermore, if the new states have a large width and can not easily be seen as a resonance, 
 their presence might be inferred by a broad increase in longitudinal mode production near the particle mass.  
 This would also be useful in probing a scenario with a heavy Higgs.
 
 Furthermore, in models of EWSB where a heavy $Z'$ mixes with the SM $Z$, there are even more dramatic
effects one could look for.  In these models the VBF aptitudes grow as $E^4$ up until the scale of the $Z'$~\cite{Cheung:2009um}. 
These models can be difficult to probe if the $Z'$ is too heavy to be produced, but one could in principle observe the unique amplitude
growth characteristic of these scenarios. 
 
Another extension of this sort of analysis using VBF as a robust probe of new physics would involve
making use of azimuthal angle correlations.  This sort of analysis has been performed with an eye toward
distinguishing the spin of the unitarizing particle~\cite{Hagiwara:2009wt}.  It would be interesting to see if this 
could yield an additional handle on background suppression or signal enhancement.

Finally, we note that while our article focuses on the $E^2$ growth in $V_LV_L$ amplitudes associated with 
new physics, for many interesting cases the rates are so low that with the LHC we can probably only hope 
to observe this growth when integrated over all higher energies.   A luminosity upgrade to the LHC~\cite{Scandale:2008zzc} 
might allow for the {\it differential} detection of this growth: given enough data,  one could bin longitudinal fractions 
according to the $\hat{s}$ of the reconstructed $VV$ system, allowing for a differential detection of the amplitude growth.  
\FIGURE[t]{
\includegraphics[scale=0.6]{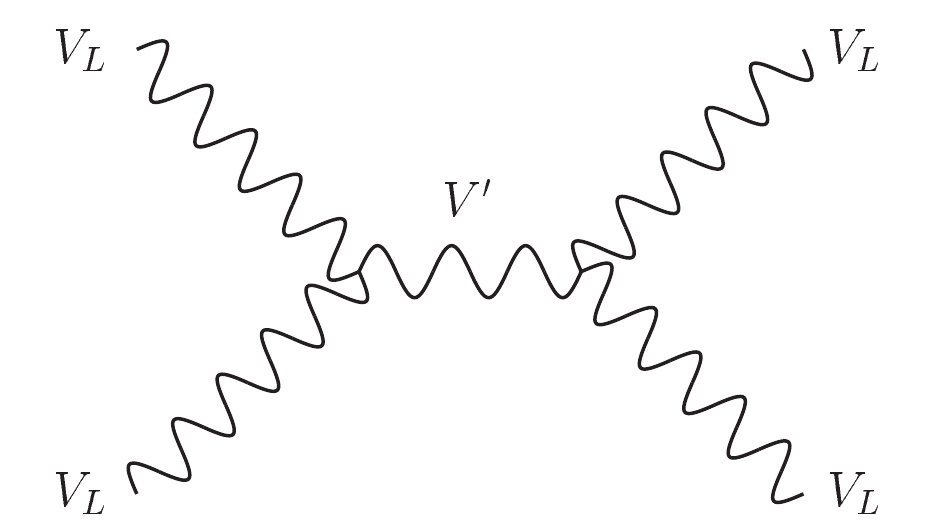} 
\label{fig:spin_one_unit} 
\caption{Unitarity restoration through a spin-1 particle.}
}
\section{Conclusions}
\label{sec:concl}
In this article we have introduced a powerful new technique for identifying signs
 of beyond the SM physics associated with the EWSB by probing VBF processes at the LHC.  

We began by motivating our decision to  study models of EWSB employing a light Higg-like
 particle with couplings deviating from those of the SM.  Theories with a light Higgs boson are
 favored by the current electroweak precision data. 
 However, this type of model is the most difficult 
 to distinguish from the SM, especially if the new physics particles are very heavy.  However, it is 
 also the scenario in which VBF can be most useful, because for such a scenario the amplitude
  for $V_LV_L\rightarrow V_LV_L$ scattering exhibits a non-SM $E^2$ growth until new physics comes into play.

Past analyses designed to measure this $E^2$ growth were reviewed and updated to account for 
the effects of the parton shower and jet clustering.  While the cuts pioneered by these works can be
 very powerful in reducing the SM background, we demonstrate that there is still a significant 
 $\mathcal{O}(100\%)$ rate uncertainty attributable to factorization scale 
 ambiguities.  Thus, we show that in the absence of higher order calculations that might give us 
 some guidance on the correct scale treatment, rate information alone may not be sufficient to distinguish 
 the signs of new physics.

We then propose our new technique, which uses the semi-leptonic decay mode of the $VV$ system 
to fully reconstruct events and obtain the decay angle distributions for the $V$ daughters.  These 
distributions can be decomposed into longitudinal and transverse components, allowing us to
 measure the $E^2$ growth in scattering amplitudes associated with new physics by looking 
 for the relative increase in longitudinal production.  We demonstrate that these results are 
 insensitive to the scale ambiguities that trouble rate measurements.

In closing, we wish to reiterate that polarization measurements of VBF final states are a 
powerful, robust probe of new physics associated with the EWSB.  
Although we have only employed them here to  
study light SM-like Higgs scenarios, they would be useful in more general scenarios of EWSB
as long as the longitudinal gauge bosons are significantly involved.
Such measurements may prove to be our best tool in understanding the physics of EWSB at the LHC.

\acknowledgments{The authors would like to acknowledge useful discussions with Johan Alwall, Jiji Fan, Kentarou Mawatari, and Matt Schwartz.  
The work of T.H. is supported in part by the DOE under grants DE-FG02-95ER40896 and 
W-31-109-Eng-38, and in part by the Wisconsin Alumni Research Foundation.  L.-T. W. was 
supported by NSF grant PHY-0756966 and DOE grant DE-FG02-90ER40542.}
\appendix

\section{Scattering Amplitudes for Longitudinal  Gauge Bosons and Partial Wave Unitarity}
\label{sec:scatamp}
For completeness, we will here review the high energy behavior of longitudinal gauge boson scattering.  
This will demonstrate why we expect the increase in the scattering amplitudes for non-SM Higgs couplings. 
It will also help us establish the partial wave unitarity bound for longitudinal gauge boson scattering, 
which is of practical importance
 for our simulation.  We will make use of the Goldstone equivalence theorem, which says that the scattering 
 behavior of the longitudinal gauge bosons is the same as that of the eaten Goldstones, up to corrections of
  order $\mathcal{O}(m_W/E)$.  Note that while we will only explicitly calculate the behavior of 
  $W_L^+W_L^-\rightarrow W_L^+W_L^-$, the other longitudinal gauge boson scattering processes are similar. 

We begin with the Lagrangian for the SM Higgs doublet with the additional dimension-6 operator we wish to study:
\be
\mathcal{L}=\frac{1}{4}\tr\left(\partial\mathcal{H}^\dagger \partial\mathcal{H}+
{\mu^2}\mathcal{H}^\dagger\mathcal{H}\right)-\frac{\lambda}{16}\tr\left(\mathcal{H}^\dagger\mathcal{H}\right)^2
+\frac{c_H}{32f^2}\left[\partial\tr\left(\mathcal{H}^\dagger\mathcal{H}\right)\partial\tr\left(\mathcal{H}^\dagger\mathcal{H}\right)\right]
\ee
where $\mathcal{H}=\phi\cdot\sigma$ for real fields $\phi_i$ ($i=0\leftrightarrow3$) and  $\sigma=(1,\vec{\sigma})$.  
The SM Higgs potential corresponds to $c_H^{}=0$.
Expanding around the minima $\langle \phi_0\rangle=\mu/\sqrt{\lambda}$ 
one finds new derivative interactions proportional to $c_H$.  Those relevant to 
$\phi_+\phi_-\rightarrow \phi_+\phi_-$ scattering at lowest order are:
\be
\mathcal{L}\supset-v\lambda h\phi_+\phi_--\frac{\lambda}{2}\phi_+^2\phi_-^2
+\frac{c_H}{2f^2}\left(\phi_+^2 (\partial\phi_-)^2+\phi_+\phi_-\partial\phi_+\partial\phi_-+2v\phi_+\partial\phi_-\partial h\right)+{\rm h.c.}
\ee
where we have denoted the shifted $\phi_0$ field by $h$ and written $\phi_{1,2}$ in terms of their charge eigenstates $\phi_\pm$.  
Also, note that there is an additional kinetic term for $h$:
\be
\mathcal{L}\supset \frac{c_H v^2}{2f^2}(\partial h)^2
\ee
so that in going to canonical normalization we must insert a factor of $N=1/\sqrt{1+c_Hv^2/f^2}$ 
for every $h$ encountered at a vertex.  The tree level amplitude becomes
\be
\label{eq:highendscat}
\mathcal{M}(\phi_+\phi_-\rightarrow\phi_+\phi_-)=-4i\lambda+i\frac{c_H}{f^2}s
-\frac{iN^2}{(s-m_h^2)}\left(2\lambda v+\frac{c_H v}{2f^2}s\right)^2+(s\leftrightarrow t)
\ee
where $m_h=\sqrt{2\lambda}v$.  Working in the limit $s,t\gg m_h^2$ we find  
\be
\label{eq:scatampanom}
\mathcal{M}(\phi_+\phi_-\rightarrow\phi_+\phi_-)\approx i\frac{c_H}{f^2}\left(1-N^2\frac{ c_H v^2}{4 f^2 }\right)\left(s+t\right)
\ee
which shows the $E^2$ growth in the amplitude that we expect.  In this limit, the $J=0$ partial wave is:
\be
a_0=\frac{1}{16\pi s}\int^0_{-s}\frac{c_H}{f^2}\left(1-N^2\frac{ c_H v^2}{4 f^2 }\right)\left(s+t\right) dt
=\frac{c_H s}{32 \pi f^2}\left(1-N^2\frac{ c_H v^2}{4 f^2 }\right)
\ee
Partial wave unitarity is violated when $| {\rm Re} (a_I) | \geq 1/2$, so the unitarity bound is saturated when
\be
s_{\rm max}=\frac{16 \pi v^2}{c_H\xi\left(1-c_H\xi N^2/{4 }\right)}
\ee
To stay clear of this limit, we limit ourselves to studying events for which $s \leq 2~{\rm TeV}$ 
(corresponding to $|c_H \xi| \leq 0.6$).

We note that one may gain further intuition into the longitudinal gauge boson system by considering the parameterization
\be
\mathcal{H}=\left(v+h\right)e^{i\vec{\pi}\cdot \vec{\sigma}/v}
\ee 
Here we have shifted our field definitions so that the $\pi$ transform non-linearly. 
 In this language, the relevant terms in the Lagrangian become
\be
\mathcal{L}\supset\frac{\sqrt{\lambda}}{\mu}h\partial \pi_+\partial\pi_-
+\frac{\lambda}{6 \mu^2}\left(\pi_+^2(\partial \pi_-)^2-\pi_+\pi_-\partial\pi_+\partial\pi_-\right)+{\rm h.c.}
\ee
As before, the kinetic term of $h$ is shifted, so we must add a factor of $N$ at every point we encounter an $h$ at a vertex. 
 Note, however, that in this case all of the operators come with $\partial \pi$ terms.  Computed in this way, 
 the amplitude $\mathcal{M}(\pi_+\pi_-\rightarrow \pi_+\pi_-)$ shows the same behavior as Eq.~(\ref{eq:highendscat}), as it must, but this is the result of 
a {\it non-cancelation} of derivatives between the four-point operator and the $h$-exchange in the $t\ \&\ s$-channels, rather than because of a new vertex.

Using these results we can compare the scattering in the Higgsless case to that of the case 
where the Higgs has anomalous couplings.  To consider the Higgsless case we 
set $c_H=0$ in Eq.~(\ref{eq:highendscat}) and consider the $\sqrt{s} \ll m_h$ limit
using $m_h=\sqrt{2\lambda}v$.  We find
\be
\sigma_{\rm no-higgs}\propto |\mathcal{M}|^2= \frac{4}{v^4}(s+t)^2
\ee 
wheras for the case of a light Higgs with anomalous couplings we find from Eq.~(\ref{eq:scatampanom})
\be
\sigma_{\rm anom-higgs}\propto|\mathcal{M}|^2= \frac{(c_H\xi)^2}{v^4}(s+t)^2
\ee
under the assumption 
\be
\label{eq:cond}
\frac{c_H\xi}{2} \gg  \frac{m_h^2}{s}
\ee
This is how we arrived at Eq.~(\ref{eq:llscat}).  For $m_h\sim100~{\rm GeV}$ this is 
true for scattering at the TeV scale as long as ${c_H\xi} \gsim 1/10$.  At lower values of the anomalous coupling the 
dominant effect comes from interference effects proportional to $s+t$ instead of $(s+t)^2$.  Thus smaller values of the 
anomalous couplings have a qualitatively different energy behavior that of larger values, making them especially
difficult to resolve.
\section{Overview of Jet Algorithms}
\label{sec:jetreview}

While a comprehensive review of jet algorithms is beyond the scope of this work (see~\cite{Salam:2009jx} for a recent review), 
here we will provide a short overview so the reader can quickly gain intuition into subjet techniques.

Jet algorithms can roughly be divided into two categories: cone algorithms, which function as cookie-cutters to 
stamp out jets from calorimeter cells, and sequential recombination algorithms, which build up a jet by merging 
four-momenta one by one in a prescribed order.  Here we will focus on recombination algorithms.

Each of these algorithms functions by defining a distance measure between every pair of four-momenta and for each four-momenta individually:
\be
\label{eq:dm}
d_{ij}=\min(p_{Ti}^{2n},p_{Tj}^{2n})\left(\frac{R_{ij}}{R_0}\right)^2,\ d_{iB}=p_{Tj}^{2n}
\ee
for jets $i$ and $j$.  If the smallest distance measure at a given stage in clustering is between two four-momenta they are merged, otherwise the four-momenta
with the smallest $d_{iB}$ is declared a jet and removed from the queue.
\TABLE{
\parbox{\textwidth}{
\begin{center}
\begin{tabular}{|c|c|c|c|}
\hline
Algortithm& $n$ &Approximate clustering order \\
\hline
$k_T$ & $1$ & soft$\rightarrow$ hard\\
C/A & $0$ & near$\rightarrow$far (in $y$-$\phi$)\\
anti-$k_T$ & $-1$ & hard$\rightarrow$ soft\\
 \hline
\end{tabular}
\end{center}
\caption{Parameterization and approximate behavior of sequential recombination jet algorithms.\label{tab:jet_clustering} }
}
}

The different sequential recombination algorithms are distinguished by value of $n$ appearing in Eq.~(\ref{eq:dm}).  These values determine the 
clustering order, whether one clusters beginning with hard four-momenta, soft four-momenta, or by angle (see Table~\ref{tab:jet_clustering} and Fig.~\ref{fig:jets}).  
For the subjet analysis at hand, where reconstructing the softer subjet is essential, we therefore use the $k_T$ algorithm which begins 
by clustering softer jets, preventing them from being cannibalized by the harder subjet.

\FIGURE{
\includegraphics[scale=0.75]{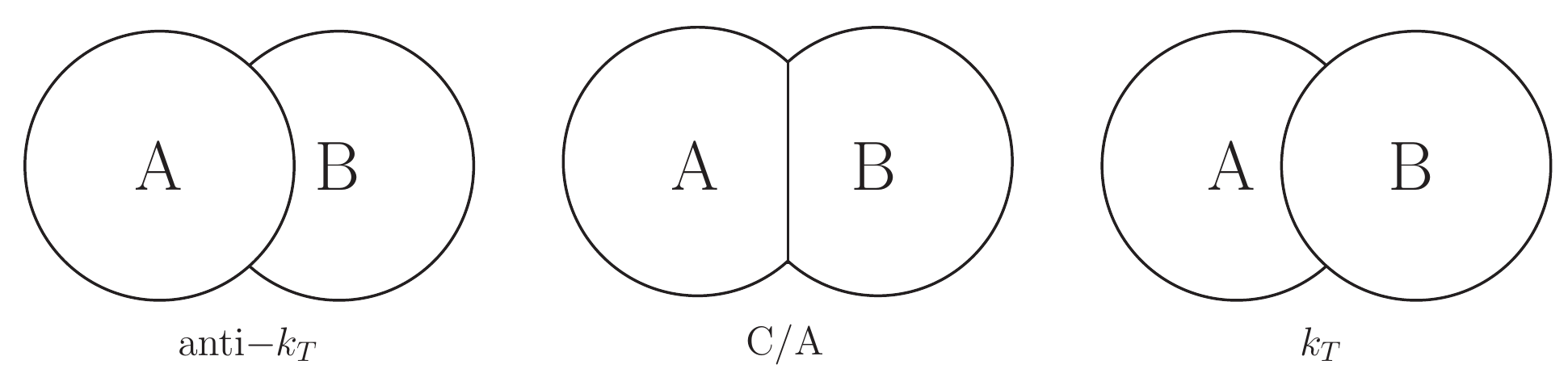} 
\caption{
\label{fig:jets} Approximate clustering behavior of jets for the different sequential recombination algorithms assuming $p_{T}^A > p_T^B$.  
Note that while we have shown the jets as being circular, the $k_T$ jets can behave in a non-circular, wandering way.}  }
\bibliography{ww}
\bibliographystyle{jhep}
\end{document}